\documentclass[sigconf,screen]{acmart} 

\usepackage{graphicx}
\usepackage[capitalise]{cleveref}
\usepackage{balance}
\usepackage[table,xcdraw]{xcolor}
\usepackage{tabularx}
\usepackage{multirow,multicol,makecell}
\usepackage{framed}
\usepackage{nomencl} 
\usepackage{xparse}
\usepackage{pgfplots}
\usepackage{pgfplotstable}
\usepackage{subcaption}
\usepackage{caption}
\usetikzlibrary{positioning,calc,matrix}
\usepackage{booktabs}
\usepackage{algpseudocode}
\usepackage{algorithm}
\usepackage[inline,shortlabels]{enumitem}
\usepackage[many]{tcolorbox} 
\usetikzlibrary{patterns} 
\usepackage{balance}
\usepackage{microtype}
\usepackage{sansmath}
\usepgfplotslibrary{groupplots} 

\pgfdeclarelayer{foreground}
\pgfsetlayers{main,foreground}

\pgfplotsset{
    tick label style = {font=\sansmath\sffamily\small},
    every axis label = {font=\sansmath\sffamily\small},
    legend style = {font=\sansmath\sffamily\small},
    label style = {font=\sansmath\sffamily\small},
    box plot/.style={
        /pgfplots/.cd,
        black,
        only marks,
        mark=-,
        mark size=\pgfkeysvalueof{/pgfplots/box plot width},
        /pgfplots/error bars/y dir=plus,
        /pgfplots/error bars/y explicit,
        /pgfplots/table/x index=\pgfkeysvalueof{/pgfplots/box plot x index},
    },
    box plot box/.style={
        /pgfplots/error bars/draw error bar/.code 2 args={%
            \draw  ##1 -- ++(\pgfkeysvalueof{/pgfplots/box plot width},0pt) |- ##2 -- ++(-\pgfkeysvalueof{/pgfplots/box plot width},0pt) |- ##1 -- cycle;
        },
        /pgfplots/table/.cd,
        y index=\pgfkeysvalueof{/pgfplots/box plot box top index},
        y error expr={
            \thisrowno{\pgfkeysvalueof{/pgfplots/box plot box bottom index}}
            - \thisrowno{\pgfkeysvalueof{/pgfplots/box plot box top index}}
        },
        /pgfplots/box plot
    },
    box plot top whisker/.style={
        /pgfplots/error bars/draw error bar/.code 2 args={%
            \pgfkeysgetvalue{/pgfplots/error bars/error mark}%
            {\pgfplotserrorbarsmark}%
            \pgfkeysgetvalue{/pgfplots/error bars/error mark options}%
            {\pgfplotserrorbarsmarkopts}%
            \path ##1 -- ##2;
        },
        /pgfplots/table/.cd,
        y index=\pgfkeysvalueof{/pgfplots/box plot whisker top index},
        y error expr={
            \thisrowno{\pgfkeysvalueof{/pgfplots/box plot box top index}}
            - \thisrowno{\pgfkeysvalueof{/pgfplots/box plot whisker top index}}
        },
        /pgfplots/box plot
    },
    box plot bottom whisker/.style={
        /pgfplots/error bars/draw error bar/.code 2 args={%
            \pgfkeysgetvalue{/pgfplots/error bars/error mark}%
            {\pgfplotserrorbarsmark}%
            \pgfkeysgetvalue{/pgfplots/error bars/error mark options}%
            {\pgfplotserrorbarsmarkopts}%
            \path ##1 -- ##2;
        },
        /pgfplots/table/.cd,
        y index=\pgfkeysvalueof{/pgfplots/box plot whisker bottom index},
        y error expr={
            \thisrowno{\pgfkeysvalueof{/pgfplots/box plot box bottom index}}
            - \thisrowno{\pgfkeysvalueof{/pgfplots/box plot whisker bottom index}}
        },
        /pgfplots/box plot
    },
    box plot median/.style={
        /pgfplots/box plot,
        /pgfplots/table/y index=\pgfkeysvalueof{/pgfplots/box plot median index}
    },
    box plot width/.initial=1em,
    box plot x index/.initial=0,
    box plot median index/.initial=1,
    box plot box top index/.initial=2,
    box plot box bottom index/.initial=3,
    box plot whisker top index/.initial=4,
    box plot whisker bottom index/.initial=5,
}

\newcommand{\boxplot}[2][]{
    \addplot [box plot median,#1] table {#2};
    \addplot [forget plot, box plot box,#1] table {#2};
    \addplot [forget plot, box plot top whisker,#1] table {#2};
    \addplot [forget plot, box plot bottom whisker,#1] table {#2};
}
\pgfplotsset{compat=1.18}

\copyrightyear{2025}
\acmYear{2025}
\acmConference[MIDDLEWARE '25]{26th International Middleware Conference}{December 15--19, 2025}{Nashville, TN, USA}
\acmBooktitle{26th International Middleware Conference (MIDDLEWARE '25), December 15--19, 2025, Nashville, TN, USA}
\acmDOI{10.1145/3721462.3770777}
\acmISBN{979-8-4007-1554-9/2025/12}

\begin{document}

\title{\textit{\textsc{Roadrunner}}: Accelerating Data Delivery to WebAssembly-Based Serverless Functions}


\author{Cynthia Marcelino}
\orcid{0000-0003-1707-3014}
\affiliation{%
  \institution{Distributed Systems Group, TU Wien}
  \city{Vienna}
  \country{Austria}}
\email{c.marcelino@dsg.tuwien.ac.at}

\author{Thomas Pusztai}
\orcid{0000-0001-9765-6310}
\affiliation{%
  \institution{Distributed Systems Group, TU Wien}
  \city{Vienna}
  \country{Austria}}
\email{t.pusztai@dsg.tuwien.ac.at}

\author{Stefan Nastic}
\orcid{0000-0003-0410-6315}
\affiliation{%
  \institution{Distributed Systems Group, TU Wien}
  \city{Vienna}
  \country{Austria}}
\email{snastic@dsg.tuwien.ac.at}

\renewcommand{\shortauthors}{Marcelino et al.}

\begin{abstract}

  Serverless computing provides infrastructure management and elastic auto-scaling, therefore reducing operational overhead. By design serverless functions are stateless, which means they typically leverage external remote services to store and exchange data. 
  Transferring data over a network typically involves serialization and deserialization. These operations usually require multiple data copies and transitions between user and kernel space, resulting in overhead from context switching and memory allocation, contributing significantly to increased latency and resource consumption.
  
  

  To address these issues, we present Roadrunner, a sidecar shim that enables near-zero copy and serialization-free data transfer between WebAssembly-based serverless functions. Roadrunner reduces the multiple copies between user space and kernel space by mapping the function memory and moving the data along a dedicated virtual data hose, bypassing the costly processes of serialization and deserialization. This approach reduces data movement overhead and context switching, achieving near-native latency performance for WebAssembly-based serverless functions. Our experimental results demonstrate that Roadrunner significantly improves the inter-function communication latency from 44\% up to 89\%, reducing the serialization overhead in 97\% of data transfer, and increasing throughput by 69 times compared to state-of-the-art WebAssembly-based serverless functions.
\end{abstract}


\maketitle

\section{Introduction}
\label{sec:intro}
Serverless functions are short-lived and stateless, which means they often rely on external remote services to store and exchange data~\cite{RisePlanetServerless2023,castro2019rise,li2022serverless,2017serverlessDesign, ServerlessInTheWild}. Since serverless functions are short-lived by design, a single function cannot be directly addressed. Therefore, clients rely on the platform ingress and Load Balancers to access the serverless function~\cite{state-of-serverless, serverlessTrends, onestep,ServerlessComputingSurvey}. 
WebAssembly (Wasm) has emerged, offering serverless functions with strong isolation, decreased cold starts, and near-native execution speed. Wasm executes small binary modules within a lightweight, memory-safe, and secure sandbox, enabling cross-platform portability while maintaining minimal overhead~\cite{faasm,LightweightFaaS,goldfish2024,sledge,cncf2023wasm,PerformanceIsolation2025}.
Wasm relies on the WebAssembly System Interface (WASI) to interact with the host for essential tasks, such as accessing network interfaces, introducing an additional overhead in wasm-based serverless functions~\cite{Cwasi2023,PushingWasm,WasmExecutionModel,bringingSpeedWasm}. 
The WASI dependency for host interactions, including system calls needed for network access, adds extra overhead in serverless functions since these functions typically rely on remote services to retrieve data. 
These operations lead to multiple context switches and data copies between user and kernel space, increasing function execution time and resource consumption. 

\begin{figure*}[t]
    \includegraphics[width=\linewidth]{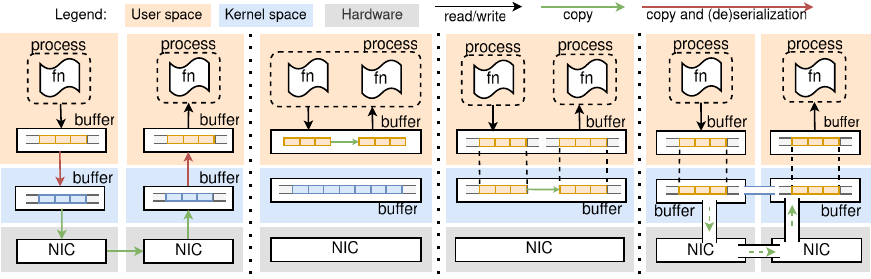}
    
    \begin{subfigure}[t]{0.29\textwidth}
        \vspace{-1.5em}
        \caption{HTTP Data Transfer} 
        \label{fig1_a}
    \end{subfigure}\hfill
    \begin{subfigure}[t]{0.2\textwidth}
        \vspace{-1.5em}
        \caption{User Space Data Transfer} 
        \label{fig1_b}
    \end{subfigure}\hfill
    \begin{subfigure}[t]{0.21\textwidth}    
        \vspace{-1.5em}
        \caption{Kernel Space Data Transfer} 
        \label{fig1_c}
    \end{subfigure}\hfill
    \begin{subfigure}[t]{0.27\textwidth}
        \vspace{-1.5em}
        \caption{Network Data Transfer} 
        \label{fig1_d}
    \end{subfigure}\hfill
\caption{Overview of standard data transfer between Serverless functions and Roadrunner communication models: a) Standard HTTP Data Passing between functions with Serialization/Deserialization, b) User Space Data Transfer, c) Kernel Space Data Transfer between Serverless functions, serialization-free, and d) Serialization-free Network Data Transfer.}
\label{fig:arch_overview}
\end{figure*}

In serverless computing, functions typically exchange data via network protocols such as HTTP, which involves serialization of the requested data at the source function and its deserialization at the target (\cref{fig1_a}). Serialization involves converting potentially complex data structures into binary sequences for transfer, followed by deserialization to reconstruct the data structures at the destination. Serialization, along with other low-level operations such as memory allocation, compression, and network stacking, accounts for up to 25\% of CPU cycles~\cite{SerialBigData2020,AtackKiller,cornflakes,Skyway2020,zero-change}. 
During a data transfer, the application in user space prepares the data for the syscall by serializing it into a byte stream. Upon entering kernel space, 
the syscall copies the byte stream buffers to kernel-space buffers. During this process, the kernel initializes new structures for the transmission, including copying the address,  control information, and validating its integrity and permissions. These multiple copies ensure data integrity and security but add overhead due to the repeated copies during context switches~\cite{ibm,ontime_recv,linuxkernel2024,breakfast,LazyMemcopy,AchievingZeroCopyRPC,BLOCK2017S66,10.1145/3158644}.
Zero-copy techniques~\cite{Serialization2024,serial-hardware,naos} such as RDMA can completely bypass Kernel and CPU, but they require specialized hardware~\cite{Serialization2024,wei2023no,copik2021rfaas}. In Wasm, serialization costs are even higher due to Wasm constraints such as single-threaded execution, which forces the processing of complex tasks to be performed sequentially~\cite{bringingSpeedWasm,webassemblyman_datatypes}. 

Although Wasm offers near-native execution latency, its strict isolation mechanisms introduce additional overhead when interacting with the host through the WASI interface, which can become a bottleneck for IO-bound workloads. To address this, we introduce Roadrunner, a novel runtime shim that enables near-zero copy, serialization-free data transfer between functions without requiring specialized hardware such as RDMA-enabled components. Roadrunner establishes a virtual data hose between functions, allowing data to be transferred from memory and bypassing costly serialization and deserialization processes typically required for inter-function communication.
Specifically, Roadrunner is tailored for data-intensive edge-cloud scenarios (e.g., ML-based image processing, traffic data analytics), which involve multiple functions that exchange ephemeral data, such as streaming ingestion, frame extraction, data processing, and ML inference. Existing state-of-the-art approaches, such as HTTP communication, either incur high overhead due to WASI-mediated communication or require Wasm runtime modification, compromising Wasm's isolation guarantees~\cite{WasmContainerRuntime2025,PushingWasm,faasm}. To address that, Roadrunner introduces novel data-passing abstractions for intra- and inter-function communication that preserve Wasm's sandboxing by ensuring that only the specific function shim has access to the shared data. In trusted, single-tenant serverless workflows, where functions from the same workflow and tenant are colocated within the same Wasm VM, Roadrunner enables low-latency, near-zero copy communication while maintaining strict isolation.
We summarize our contributions as follows:

\begin{itemize}
\item \textit{Roadrunner: A novel serverless communication middleware} that enables efficient data transfer between Wasm-based serverless functions across user space, kernel space, and the network during runtime. User-space communication (\cref{fig1_b}) operates directly on Wasm linear memory for the Wasm VM with multiple Wasm modules. Kernel-space communication (\cref{fig1_c}) enables co-located functions to exchange data via host mechanisms, skipping network transfer. Network communication (\cref{fig1_d}) establishes lightweight data transfer for inter-node functions. This architecture achieves low latency, therefore maximizing the function performance for both local and remote function data exchange. 

\item \textit{A near-zero copy mechanism for inter-function communication} that eliminates multiple data copies between user space and kernel space without requiring specialized hardware such as RDMA-enabled components. 
Roadrunner establishes a virtual data hose that allows data written to it to prompt the kernel to allocate memory buffers and retain them in its address space. When a read operation occurs, Roadrunner leverages the kernel to reuse the same memory pages for the target function instead of copying the data, thereby avoiding duplication. It reduces intermediate memory copies, minimizes resource overhead, and ensures efficient data transfer between functions. Roadrunner improves the latency by 44\% to 89\%, increasing the throughput by 69 times compared to the state-of-the-art Wasm runtime.

\item \textit{A serialization-free data transfer mechanism that} operates directly on Wasm VM linear memory and manages memory allocation dynamically. Roadrunner allocates memory in the target function to store the incoming data and accesses the memory region specified by the source function using direct memory pointers. The data is transferred by referencing these memory regions without intermediate serialization or deserialization. By directly handling memory allocation and data transfer at the raw binary level, Roadrunner eliminates unnecessary data copies and context switches. Roadrunner reduces the serialization impact by 97\% during data transfer between wasm serverless functions.

\end{itemize}

\pgfplotstableread{
Label    Pull    Unpacking Runtime Cold
helloRC  914.6    977.6    0.5916  1.8922
helloW   34       960      0.260    0.994
resizeC  915.8    999.4    0.631    1.9152
resizeW  41       978.8    4.651   1.0198
}\motivation

\pgfplotstableread{
Label                 Transfer    Serialization TransferNorm SerializationNorm
1MB-Container  0.014581586 0.002361838   86.06 13.94
1MB-Wasmedge   0.017468757 0.013808238   55.85 44.15
60MB-Container  0.593202801 0.073222246   89.01 10.99
60MB-Wasmedge   0.632984033 1.084127719   36.86 63.14
100MB-Container 1.081091731 0.096302649   91.79 8.21
100MB-Wasmedge  1.295975991 1.555044937   45.45 54.55
}\data

\begin{figure}[!t]
\hspace{-3em}
\begin{subfigure}[t]{0.48\linewidth}
\begin{tikzpicture}
    \begin{axis}[
        ybar stacked,
        bar width=12pt,
        ylabel={Latency (sec)},
        ylabel style={yshift=-5pt,font=\footnotesize}, 
        ymajorgrids=true,
        grid style=dashed,
        height=3.8cm,
        width=4cm,
        xtick=data,
        ymax=7.5,
        symbolic x coords={helloRC, helloW, ,resizeC, resizeW},
        xticklabels={
            {Cont}, {Wasm},
            {Cont}, {Wasm}
        }, 
        xticklabel style={
            align=center, 
            font=\tiny 
        },
        group style={
            group size=2 by 1, 
            horizontal sep=15pt, 
            xlabels at=edge bottom
        },
        legend style={at={(0.5,1.24)}, anchor=north, draw=none,legend columns=-1}, 
        enlarge x limits=0.15 
    ]
    
        \addplot [fill=yellow!60, postaction={pattern=dots}] 
            table [x=Label, y=Cold] {\motivation};

        \addplot [fill=green!80!black, postaction={pattern=crosshatch}] 
            table [x=Label, y=Runtime] {\motivation};    

        \node at (axis cs:helloRC, 3.3) [anchor=center,align=center, font=\scriptsize\color{blue}] {76.8\\MB};
        \node at (axis cs:helloW, 2) [anchor=center, align=center, font=\scriptsize\color{blue}] {47.8\\KB};

        \node at (axis cs:resizeC, 3.3) [anchor=center, align=center, font=\scriptsize\color{blue}] {76.9\\MB};
        \node at (axis cs:resizeW, 6.4) [anchor=center, align=center, font=\scriptsize\color{blue}] {3.19\\MB};

        \legend{Cold Start, Execution}
    \end{axis}
    \node at (2.6cm, 1.1cm) [anchor=center, rotate=90, font=\footnotesize\color{blue}] {Image Size};
    \node at ([yshift=-2pt,xshift=5pt] current bounding box.south) [align=center, font=\tiny] {
        Hello World \quad \quad  Resize Image
    };
\end{tikzpicture}
\caption{}
\label{fig:coldstarts}
\end{subfigure}
\hspace{-2em}
\begin{subfigure}[t]{0.48\linewidth}
    \begin{tikzpicture}
    \begin{axis}[
        ybar stacked,
        ymin=0,
        ymax=110,
        bar width=12pt,
        ylabel={Normalized Latency (\%)},
        ylabel style={yshift=-5pt,font=\footnotesize}, 
        ymajorgrids=true,
        grid style=dashed,
        height=3.8cm,
        width=5.5cm,
        xtick=data,
        symbolic x coords={1MB-Container, 1MB-Wasmedge,,60MB-Container, 60MB-Wasmedge,,100MB-Container, 100MB-Wasmedge},
        xticklabels={
            {Cont}, {Wasm},
            {Cont}, {Wasm},
            {Cont}, {Wasm}
        }, 
        xticklabel style={
            align=center, 
            font=\tiny 
        },
        group style={
            group size=2 by 1, 
            horizontal sep=15pt, 
            xlabels at=edge bottom
        },
        legend style={at={(0.55,1.24)}, anchor=north, draw=none,legend columns=-1}, 
    ]
        \addplot [fill=orange!80, postaction={pattern=north east lines}] 
            table [x=Label, y=TransferNorm] {\data};

        \addplot [fill=blue!60, postaction={pattern=north west lines}] 
            table [x=Label, y=SerializationNorm] {\data};

        \legend{Transfer, Serialization}
    \end{axis}
    \node at ([yshift=-2pt,xshift=14pt] current bounding box.south) [align=center, font=\tiny] {
        1MB \quad \quad \quad \quad \quad \quad 60MB \quad \quad \quad \quad \quad 100MB
    };
\end{tikzpicture}
\caption{}
\label{fig:serialization}
\end{subfigure}
\caption{Cold start and execution latency (a), and normalized I/O breakdown (b) for functions using Docker Containers (Cont) and Wasm. (a) shows Wasm lowers cold starts for both ``Hello World'' (no WASI) and ``Resize Image'' (with WASI), though WASI increases execution time. (b) compares transfer and serialization overhead across various input sizes.}
\label{fig:motivation}
\end{figure}
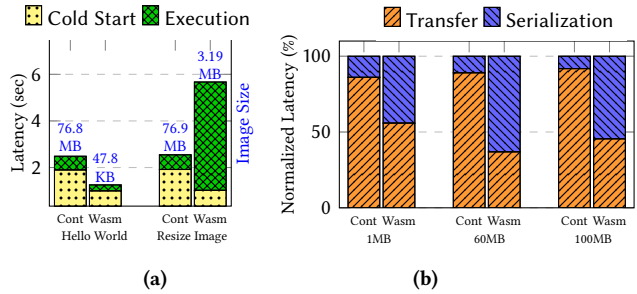

\section{Background and Motivation}
\label{sec:motiv}


\subsection{WebAssembly in Serverless Computing}



\paragraph{Portability.} Wasm is a portable, low-level binary instruction format designed as a compilation target for high-level languages such as Rust, C, C++, and Go. Wasm modules are platform-agnostic and can be executed on different architectures (e.g., x86, arm, and aarch64)~\cite{BidirectionalServerless, Wasmbpf2024,webAssemblyCommonLayer}. 
Wasm portability contrasts with traditional serverless platforms, which execute functions within OCI bundles (containers) using runtimes such as RunC~\cite{runc} and Crun~\cite{crun}, which rely on the host kernel and Linux primitives such as \texttt{cgroups} and \texttt{namespaces} for isolation~\cite{man7_cgroups,man7_namespaces,fnAsFunction,copik2024praas,parola2024sure}, running directly on the host OS. 
In Wasm, the function code is compiled into small binaries and executed in a secure, sandboxed environment (i.e., Wasm VM) and, therefore, more lightweight compared to Docker images, as they do not require a base image operating system (OS) or extensive initialization provision~\cite{WasmExecutionModel,PushingWasm,webAssemblyCommonLayer}. 

\emph{Linear Memory.} Wasm uses a linear memory model, where memory is represented as a contiguous, byte-addressable array. The allocated memory for the Wasm VM can dynamically grow, providing isolated and efficient access for computation between Wasm modules, enabling safe memory management and isolation between the module and host environment~\cite{cncf2023wasm, BidirectionalServerless, bringingSpeedWasm}.

\emph{WASI Overhead.} Wasm follows the deny-by-default principle, which means Wasm binaries do not have access to the host machine unless explicitly requested. To access host functionalities such as the file system and network interfaces, Wasm leverages WASI~\cite{faasm,sledge,PushingWasm,goldfish2024}. 
\cref{fig:coldstarts} presents the effect of WASI-based host communication on function execution time. Wasm functions exhibit reduced cold start latency, with Wasm binaries of 3.19MB, whereas Docker images have approximately 77 MB. During execution, functions that do not require WASI access achieve lower latency than their Docker container counterparts. In contrast, functions interacting with the host file system through WASI exhibit increased execution times.

\subsection{Data Passing between Serverless Functions}
To exchange data, serverless functions typically rely on external remote services. 
While using remote services decouples function computation from I/O operations, allowing high scalability, it also introduces additional overhead~\cite{TheGapResearchRealWorld,serverless-why-when-how,WhereWeAreLiesAhead,whatServerlessIsAndWhatShouldBecome}. 
The most common approaches to exchange data between serverless functions are: 
\begin{enumerate*} [label=(\alph*)]
    \item \textit{Remote services} such as object storage~\cite{s3}, KVS~\cite{anna, pocket,faasTrack}, cache~\cite{InfiniCache,faast, duo,f3}, distributed storage~\cite{gilder,sonic},  are commonly used to enable direct communication between serverless functions. However, remote services might increase latency and network overhead.
    \item \textit{Direct Communication} such as message queues~\cite{sand,copik2024praas}, TCP-punch hole~\cite{fmi},  disk storage~\cite{sonic}, and local cache~\cite{ofc,sand, cloudburstSF, FaasFlow, crucial} allows serverless functions to leverage host mechanisms to exchange data, decreasing latency. Nevertheless, these approaches decrease the function isolation~\cite{Cwasi2023,pipeDevice}.
    \item \textit{Zero-copy communication} such as shared memory~\cite{faabric, nightcore, parola2024sure,faastlane} enables co-located functions to exchange data from a specific memory region. Nevertheless, it increases development complexity to ensure that only allowed functions can access the data in the shared memory.   RDMA~\cite{freeflow,naos,AchievingZeroCopyRPC,wei2023no,copik2021rfaas} enables direct memory access between nodes over the network by skipping CPU and kernel, and directly processing the data by using Network Interface Cards (NICs) to read or write directly to the memory regions specified by functions. However, RDMA usage requires specialized hardware.
\end{enumerate*}
Moreover, existing approaches such as ~\cite{sand,sonic,nightcore,faasm} rely on their own scheduling mechanisms to colocate functions to enable leveraging their data passing optimization, and local Inter Process Communication (IPC) syscalls. Roadrunner optimizes communication regardless of the scheduler's decisions.

\emph{Wasm-based Serverless Functions.} Typically, serverless functions rely on remote services over protocols such as HTTP, which involves serialization. 
Although the Wasm design enforces strict isolation, the WASI dependency for host interactions introduces overhead, particularly during data serialization~\cite{WasmContainerRuntime2025,Cwasi2023,Lumos2025}. In Wasm, serialization involves converting complex data structures within the Wasm VM into a linear, standardized format, allocating memory for the serialized output, and copying the data across the Wasm VM boundary, leading to increased execution latency and higher CPU and memory overhead.
We further investigate the WASI overhead in \cref{fig:serialization}, analyzing the impact of serialization on data transfer in the state-of-the-art serverless Docker container runtime and the Wasm runtime. In our preliminary experiments, we observed that serialization accounts for up to 15\% of the execution time of data transfer in Docker runtime, while in Wasm, it contributes up to 60\%, significantly impacting the performance of Wasm-based serverless functions.
This challenge becomes even more significant for functions that require frequent data access. Therefore, optimizing the transfer of data between functions is crucial to reduce execution time.



\begin{table*}[t]
\caption{Roadrunner Data Access Function APIs}
\label{tab:functions}
\resizebox{\textwidth}{!}{%
\begin{tabular}{@{}llll@{}}
\toprule
\textbf{Function}                        & \textbf{Description}                                                  & \textbf{Category}              & \textbf{Location} \\ \midrule

\texttt{void allocate\_memory(len)}      & Allocates a specified amount of linear memory in the Wasm VM          & Memory Management              & Function          \\

\texttt{void deallocate\_memory(address)}      & Deallocates a specified amount of linear memory in the Wasm VM          & Memory Management              & Function          \\

\texttt{void read\_memory\_wasm(address, len)}  & Reads data from a specified memory address and length in the Wasm VM  & Data Management                    & Function          \\
\texttt{(int,int) locate\_memory\_region(data)} & Returns the memory pointer and length of the specified data          & Data Management                    & Function          \\
\texttt{void send\_to\_host(address, len)} & Transfers data memory information to the host interface using the address and length & Data Management          & Function          \\ 
\texttt{void read\_memory\_host(address, len)} & Reads the data from the Wasm VM memory using a specified address and length & Data Management          & Shim    \\ 
\texttt{void write\_memory\_host(data[],address)} & Write the data into Wasm VM & Data Management          & Shim    \\ \bottomrule
\end{tabular}%
}
\end{table*}

\section{Roadrunner Data Access Model \& System Design }
\label{sec:arch}


\subsection{Roadrunner Data Access Model} \label{subsec:data_access}

\paragraph{Overview.} Roadrunner is designed to optimize Serverless inter-function communication and data passing. Roadrunner supports three communication modes: User space, Kernel Space, and Network, which are integrated with existing serverless platforms through a lightweight shim that runs beside each function. Roadrunner relies on components of serverless platforms that enable the deployment, execution, and observability of a Serverless workflow. 

\begin{figure}[t]
    \centering
    \includegraphics[width=0.9\linewidth]{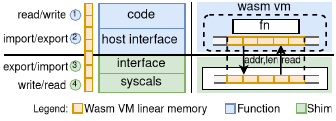}
    \caption{Roadrunner data access structured layers for data interactions on the left side and linear memory mapping of Wasm VM for addressable data access on the right side.}
    \label{fig:dataaccess}
\end{figure}

\emph{Wasm-Based Data Access.} Roadrunner enables data access within the Wasm VM by leveraging a combination of direct memory access and controlled memory management functions shown in \cref{fig:dataaccess}. Within the Wasm VM, linear memory is exposed as a contiguous block of memory and accessible through specific offsets to the host. Roadrunner utilizes this memory model to access the data stored directly in the Wasm VM sandbox. When the host needs to access data, it leverages memory pointers provided by the Wasm module to locate specific memory regions without breaking the isolation of the Wasm VM. 

\emph{Complex Data Types.} Wasm supports only primitive types (i32, i64, f32, f64), making it highly optimized for low-level operations but requiring additional mechanisms to handle complex data types such as strings. Languages such as C and Rust align well with this model, allowing direct memory management. In contrast, high-level languages such as Java and Python require embedded language runtimes within the Wasm module to manage memory, preventing direct access to linear memory and thus limiting compatibility with Roadrunner~\cite{EmpiricalWasm,EverythingOldWasm,bringingSpeedWasm, wasabi,sletten2021webassembly}. Therefore, the APIs listed in \cref{tab:functions} are designed for low-level languages to interact with Wasm memory directly and efficiently represent complex data structures such as strings. By directly using pointers to a linear memory region, we can achieve a serialization-free copy.  This serialization-free approach is valid under the assumption that all systems use the same endianness, e.g., the little-endian format found in x86 and ARM~\cite{arm_endianness}, which dominate edge and cloud environments. Furthermore, Wasm explicitly defines integer sizes (i32, i64), ensuring consistent behavior on 32-bit and 64-bit CPUs and allowing Roadrunner’s pointer-based, serialization-free access to remain valid across different architectures, as long as endianness is consistent.

\emph{Near-Zero Copy Data Exchange.} When data is passed between the shim and the Wasm VM, the shim writes or reads directly from specific memory regions, bypassing traditional intermediary steps such as file I/O or network communication.
Roadrunner coordinates memory allocation for the data and data transfer through the Wasm runtime memory interface. The data is written or read directly in the allocated memory regions. This ensures that data never leaves the Wasm VM unnecessarily, reducing the need for intermediate serialization or external storage. This approach minimizes overhead while maintaining the security and isolation properties intrinsic to Wasm.

\emph{Shared Memory.} While Roadrunner enables shared memory access between Wasm modules, it enforces strict isolation through the shim. Functions never directly access shared memory; instead, all memory operations are mediated by the shim, which manages allocation, access control, and memory mapping. Only functions of the same workflow and tenant are instantiated in the same Wasm VM, which has control over each function's linear memory space. The shim guarantees that memory visibility is limited to registered functions in the same execution context. To prevent unauthorized access and cross-tenant interference, Roadrunner restricts shim-to-Wasm access to pre-registered memory regions and applies bounds checking before any read or write operation.

\subsection{Roadrunner System Design}

\subsubsection{Lightweight} Roadrunner uses the compact binary format of Wasm to reduce overhead during function execution. Wasm modules are considerably smaller and more lightweight than traditional container images, such as those used with Docker, which usually require a base image OS and extensive initialization processes, leading to longer cold starts. This design enables quick execution while ensuring compatibility with existing serverless platforms, maintaining low overhead and high performance.

\subsubsection{Function Isolation} Roadrunner encapsulates each Wasm VM in an OCI-compliant runtime bundle, enabling interoperability with container runtime managers such as containerd. The shim runs as a sidecar alongside each function and manages the Wasm VM lifecycle, including memory configuration, binary loading, and runtime interaction. It handles all function ingress and egress, making data available to the functions via predefined APIs that expose input and output operations, allowing seamless integration with state-of-the-art orchestrators (e.g., Kubernetes) and serverless platforms (e.g., Knative, OpenFaaS, and OpenWhisk) without modifying scheduling, logging, or networking layers.

\subsubsection{Inter-function communication} Roadrunner improves communication between serverless functions using near-zero copy and serialization-free data transfer mechanisms. It achieves this through direct memory mapping and system-level operations, which reduce the number of context switches between user space and kernel space. Roadrunner supports three modes of data exchange: user space (\cref{fig1_b}), kernel space (\cref{fig1_c}), and network-based communication (\cref{fig1_d}). These modes allow for optimized data transfers based on the proximity of functions, reducing dependency on remote services and minimizing unnecessary network overhead. By enabling direct communication, Roadrunner decreases latency, supports complex function workflows, and ensures high throughput for applications that require low latency.

\subsubsection{Scalability} Roadrunner is a shim-based sidecar that lives alongside serverless functions, allowing the container orchestration tool to manage scalability. This design enables Roadrunner to optimize the data transfer while relying on the orchestrator to handle resource allocation, function placement, and horizontal scaling. Thus, Roadrunner seamlessly integrates with these platforms without adding extra complexity to scaling. The sidecar shim approach keeps Roadrunner lightweight and interoperable, allowing the orchestrator to manage increased workloads while maintaining overall system performance efficiently.

\subsubsection{Function Deployment and Lifecycle}


The Roadrunner shim plays a crucial role in managing the lifecycle of Wasm functions. During initialization, the shim creates a dedicated Wasm VM and configures the Wasm runtime, which includes setting resource limits such as memory. The function binary is then loaded into the isolated memory space of the Wasm VM. 
To ensure compatibility with standard serverless platforms, the shim packages the Wasm VM as an OCI-compliant bundle. This allows it to be executed as a container by high-level container managers such as containerd, facilitating seamless integration between orchestrators (such as Kubernetes) and Roadrunner. Consequently, Wasm-based functions can utilize the advantages of Roadrunner while using standard container orchestration tools for deployment and execution. This integration simplifies the complexities of Wasm execution and takes advantage of the existing serverless infrastructure.

\section{Roadrunner Data Transfer Mechanisms}
\label{sec:mech}
In this section, we introduce the data transfer mechanisms employed by Roadrunner, including user-space, kernel-space, and network-based communication mechanisms. 


\subsection{User Space Data Transfer}
\begin{figure}[t]
\begin{subfigure}[t]{0.44\linewidth}
\centering
    \includegraphics[width=\linewidth]{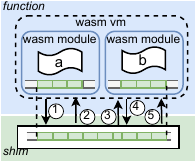}
    \caption{}
    \label{fig:user_space_datapass}
\end{subfigure}
\begin{subfigure}[t]{0.55\linewidth}
    \centering
    \includegraphics[width=\linewidth]{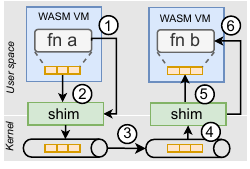}
    \caption{}
    \label{fig:kernel_mode}
\end{subfigure} 
\caption{a) User-space memory-based data transfer between Wasm serverless functions (modules) using Roadrunner's APIs for allocation, reading, and writing into Wasm linear memory. b) Kernel Space Data transfer where the shim leverages IPC to enable serialization-free data transfer.}
\end{figure}

In the user-space data transfer between functions,  shown in ~\cref{fig:user_space_datapass}, functions share the same isolation sandbox, which allows them to be executed within a single process. By sharing the same address space, functions can directly access each other's memory without requiring additional memory copies. This eliminates operations such as duplicating data between different memory regions and context switches between user space and kernel space. As a result, memory access is faster, leading to lower latency and higher throughput in the communication between functions.
Roadrunner leverages the shim to create and start the Wasm VM and manage the memory access, which enables read/write data in the function. User-mode communication requires explicit trust; thus users must specify functions within a workflow, which the shim validates using the containerd snapshot. Then, the shim initializes the VM and loads both, \textit{function a} and \textit{function b}, as Wasm modules.  When \textit{function a} needs to send data to \textit{function b}, \textcircled{\small{1}} \textit{function a} first calls \texttt{locate\_memory\_region(data)} to get the memory pointer and size of the data and forwards them to the shim. In \textcircled{\small{2}}, the shim reads it with \texttt{read\_output(address, len)} via Wasm VM memory APIs.  Next, in \textcircled{\small{3}}, after reading the data, the shim calls \texttt{allocate\_memory(len)} to reserve memory in the \textit{function b} for the incoming data. In \textcircled{\small{4}}, the shim receives the address of the allocated memory. Finally, in \textcircled{\small{5}}, the shim calls \texttt{write\_output(address, len)} in \textit{function b} with the incoming data from \textit{function a}.
This process enables efficient and secure data exchange between the two functions within the same Wasm VM, leveraging the shim to abstract memory management and streamline inter-function communication keeping the data movement within the user space.

\subsection{Kernel Space Data Transfer}

In the kernel-space data transfer mechanism, shown in \cref{fig:kernel_mode}, functions that are co-located on the same host but in different sandboxes, i.e., each function has its own dedicated shim, and they exchange data via a kernel buffer, avoiding network transfers.
Roadrunner enables quick data passing between functions by leveraging host mechanisms that allow data from one sandbox to be transferred directly via a dedicated pipe without serialization and context-switching. 

The data transfer process begins in \textcircled{\small{1}}, when \textit{function a} invokes the Roadrunner API \texttt{locate\_memory\_region(data)} to determine the memory pointer and size of the data it wants to send. This information is forwarded to the shim managing \textit{function a}. Next in  \textcircled{\small{2}}, the shim reads the data using \texttt{read\_output(address, len)} from \textit{function a}’s memory. Then, the shim leverages IPC mechanisms to send the data to the shim managing \textit{function b}, in \textcircled{\small{3}}. Upon receiving the data in \textcircled{\small{4}}, the shim for \textit{function b} calls \texttt{allocate\_memory(len)} to reserve memory within \textit{function b}’s memory space, in in \textcircled{\small{5}}.  Finally, the shim writes the data to \textit{function b}’s memory using \texttt{write\_output(address, len)} in \textcircled{\small{6}}. This kernel-mediated process ensures secure and efficient data transfer between different Wasm VM instances through IPC, leveraging the shim as a bridge between the user-space functions and the kernel.

\subsection{Network Data Transfer}

Roadrunner introduces serialization-free mechanisms that enable remote functions to pass data via a dedicated network buffer. Similar to the kernel buffer, Roadrunner creates a virtual data hose between the two hosts where the data in the user space can be directly mapped into this dedicated virtual data hose. Although the application data is located in user space, the Kernel maps it directly from the user space into the virtual data hose, thus avoiding serialization and unnecessary copies. Unlike zero-copy methods such as RDMA, which completely skips the kernel and CPU, the data passing is processed by the kernel, which means the CPU is responsible for the data movement tasks, but the data is mapped directly from the user space memory. Hence, there is no copying between the user and kernel space.

\begin{figure}[t]
    \centering
    \includegraphics[width=0.5\linewidth]{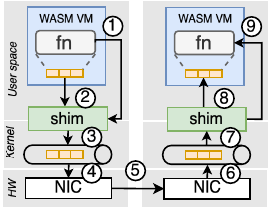}
    \caption{Network-based data transfer with near-zero copy mechanisms, mapping user-space memory into kernel buffers to optimize throughput and minimize latency.}
    \label{fig:network_data_pass}
\end{figure}

\begin{algorithm}[t]
\caption{Roadrunner Network Data Transfer}
\label{alg:networkdatatransfer}
\begin{algorithmic}[1]
\Function{FunctionA}{input, target}
    \State $ \text{(address, length)} \gets \Call{locate\_memory\_region}{\text{input}}$
    \State \Call{send\_to\_host}{address, length}
\EndFunction
\State~  
\Function{network\_data\_transfer\_source}{target}
    \State $ \text{data} \gets \Call{read\_memory\_host}{\text{address, length}}$
    \State $ \text{vdh} \gets \Call{create\_virtual\_data\_hose()}{}$ 
    \State \Call{vmsplice}{vdh, address, length}
    \State $ \text{socket\_target} \gets \Call{create\_socket}{\text{target}}$ 
    \State \Call{splice}{vdh, socket\_target, length}
    \State \Call{close\_all()}{}
\EndFunction
\State~ 
\Function{FunctionB}{}
    \State \Call{allocate\_memory}{len}
    \State \Call{deallocate\_memory}{address}
    \State $ \text{input} \gets \Call{read\_memory\_wasm}{\text{address, length}}$
\EndFunction
\State~
\Function{network\_data\_transfer\_target}{}
    \State $ \text{target\_vdh} \gets \Call{create\_virtual\_data\_hose()}{}$
    \State \Call{splice}{socket\_fd, target\_vdh, length}
    \State $\text{target\_memory\_address} \gets \Call{allocate\_memory}{\text{length}}$
    \State \Call{vmsplice}{target\_vdh, target\_memory\_address, length}
    \State \Call{write\_memory\_host}{data, target\_memory\_address}
    \State \Call{close\_all()}{}
    \State \Call{deallocate\_memory}{target\_memory\_address}
\EndFunction
\end{algorithmic}
\end{algorithm}

In the network transfer process, the shim in the source Wasm VM first retrieves data from the Wasm VM's memory space and prepares it for transmission. It begins with \textcircled{\small{1}}, where the shim in the source Wasm VM retrieves data from the memory space of the running function using the WasmEdge runtime APIs. In \textcircled{\small{2}}, the shim prepares the data and transfers it to the kernel using a syscall for near-zero copy data movement. Next, in \textcircled{\small{3}}, the kernel buffers the data and directs it towards the network stack. At this stage, in \textcircled{\small{4}}, the kernel passes the data to the network interface card (NIC), which handles the physical transmission over the network in \textcircled{\small{5}}.

On the target node, the process continues with \textcircled{\small{6}}, where the NIC receives the data and places it into the kernel buffer. The kernel then prepares this data for further processing. In \textcircled{\small{7}}, the shim on the target host uses a syscall to move the data from the kernel buffer back to user space with near-zero copy efficiency in \textcircled{\small{8}}. Finally, the shim invokes the WasmEdge APIs to allocate and write the data into the memory of the target Wasm VM in \textcircled{\small{9}}, completing the transfer. The data is now available for the function in the target Wasm VM to process. 
\cref{alg:networkdatatransfer} provides a pseudo-code representation of these steps, offering an abstraction for implementation.

\section{Roadrunner Implementation} \label{sec:impl}
Roadrunner is published as an open-source framework part of the Polaris project. It is implemented in Rust and available on Github\footnote{\url{https://github.com/polaris-slo-cloud/roadrunner}}. Roadrunner follows OCI standards, which means it can be used in any of the state-of-the-art serverless platforms. 

For user-space data transfer, we utilize the WasmEdge runtime APIs to directly access the linear memory space of Wasm modules. This allows efficient read and write operations within the same process, avoiding unnecessary data copies and reducing overhead.

For kernel-space data transfer, we employ Unix sockets as IPC mechanism. They provide a reliable and low-latency communication channel between processes, leveraging the kernel to handle data transfer efficiently. We utilize Unix sockets to enable seamless data exchange between different serverless functions while maintaining isolation and security.

For network-based data transfer, we utilize the \texttt{splice}~\cite{splice_manpage} and \texttt{vmsplice}~\cite{vmsplice_manpage} syscalls. These Linux-specific calls enable zero-copy data transfer between file descriptors, significantly reducing the overhead of data movement across network boundaries. Since \texttt{splice} and \texttt{vmsplice} bypass the user-space buffer, they allow direct data transfer between kernel buffers and function memory, improving throughput and reducing CPU utilization during network communication.

\section{Evaluation}
\label{sec:exp}

To assess the performance, scalability, and resource efficiency of Roadrunner, we design a series of experiments focusing on data-intensive serverless functions. The evaluation aims to determine how well Roadrunner optimizes inter-function communication compared to state-of-the-art runtimes. As Roadrunner introduces measurements for intra-node, i.e., user and kernel space data transfer, and inter-node communication, i.e., network data transfer, we present experiments evaluating its performance in both scenarios.

\subsection{Overview}

We perform two experiments to evaluate Roadrunner, using chained serverless workflows consisting of two I/O-bound functions, \(a\) and \(b\), which exchange serialized strings. The first experiment varies payload sizes (1MB–500MB) to assess communication overhead during sequential execution, while the second evaluates scalability through parallel workflows. These payloads reflect structured data commonly exchanged between serverless functions, with sizes representing edge–cloud serverless workflows~\cite{state-of-serverless}. In contrast, the second set assesses scalability through parallel executions. These workflows reflect real-world serverless invocation patterns, such as sequential, fan-out, and fan-in as outlined in~\cite{CloudProgrammingSimplified}. RoadRunner does not aim to provide better performance than serverless state-of-the-art solutions, such as ~\cite{sand,copik2024praas}, as these solutions operate directly on the host kernel while Roadrunner suffers from Wasm VM IO overhead (shown in \cref{fig:comparison-latency} and discussed in \cref{exp:performance}). Thus, we compare against RunC (container) as an upper bound for performance, representing the best achievable performance with Wasm. Therefore, we compare Roadrunner with state-of-the-art virtualization runtimes: RunC and Wasmedge. Since Roadrunner aims to optimize data communication, we measure the latency from the moment the source function sends data until the target function receives it.

\paragraph{Metrics} Data transfer is crucial to ensure efficient serverless, naturally stateless functions. It is critical to enable seamless communication at scale with minimal latency while avoiding exhausting computing resources. To evaluate the performance, scalability, and resource impact of Roadrunner, we conduct a series of experiments and collect results for the following metrics:

\begin{enumerate}[leftmargin=*,wide=10pt,label=\alph*)]
    \item \textit{Latency}: This metric measures the duration from when function $a$ initiates the data transfer to when function $b$ has successfully received the message. Latency is recorded in seconds to provide a clear understanding of communication delays.
    \item \textit{Throughput}: This metric evaluates Roadrunner's ability to handle high workloads without compromising performance. Throughput is measured in requests per second (RPS). For operations completed in less than one second, we extrapolate throughput by calculating the rate of requests over one second. For example, if ten requests are processed in under a second, we estimate the throughput based on their collective execution rate.
    \item \textit{Resource Usage}: For this metric, we monitor the host system's resource utilization for each function. Specifically, we track CPU usage (in percent) and memory consumption (in MB). We perform fine-grained measurements directly from the \texttt{cgroup}, enabling us to accurately capture the total CPU usage for each sandbox (container), including detailed breakdowns of user space and kernel CPU consumption. This metric allows us to compare the resource overhead introduced by Roadrunner against baseline metrics, ensuring that any increase remains minimal and within acceptable limits.
\end{enumerate}

\subsection{Experimental Setup}
Our experiments focus on evaluating chained serverless functions in a distributed environment. To execute these experiments, we employ RunC~\cite{runc} as the low-level container runtime, containerd as the high-level container manager, and WasmEdge~\cite{wasmedge} as the Wasm runtime. We use multiple nodes to facilitate remote data passing between functions. We executed the experiments 10 times and calculated the mean results. We deploy two nodes to enable remote data exchange between functions, emulating an edge–cloud setup. To approximate constrained edge–cloud conditions, we limit the available bandwidth to 100 Mbps using Linux traffic control (tc), and observe a stable round-trip latency of ~1 ms between nodes. Each node is a VM with a 4-core 2GHz vCPU, 8GB of RAM, running Ubuntu 22.04 LTS. The underlying server is running an Intel Xeon processor of the Skylake generation. The experimental workflows are implemented in Rust and leverage Wasmedge SDK for WASI-related tasks and tokio~\cite{tokio} for HTTP requests.

\subsection{Performance}\label{exp:performance}

\textbf{\underline{Intra-node:}} \textbf{\textit{Roadrunner (User space)} reduces the latency by 44\% to 89\% compared to Wasmedge and by 10\% to 80\% compared to RunC.
\textit{Roadrunner (Kernel space)} reduces the latency by 76\% to 83\% compared to Wasmedge and up to 13\% compared to RunC.}

\cref{fig:intra-different-payload} shows the performance metrics, including Total and Serialization Latency for different payload sizes across three runtimes: Roadrunner, and Wasmedge. \cref{fig:intra-total_latency-performance} and \cref{fig:intra-total_throughput-performance} represent the total latency and throughput, respectively, including data transfer and data serialization times. Roadrunner (User space) achieves the lowest latency due to near-zero copy transfers, followed by Roadrunner (Kernel space). RunC and WasmEdge exhibit significantly higher latencies.\cref{fig:intra-serial_latency_performance} and \cref{fig:_intra-serial_throughput-performance} show exclusively the serialization impact on the function data transfer. Roadrunner (User space) and RunC demonstrate negligible latency. Here, it is possible to notice the serialization impact on the Wasmedge function, which is responsible for the high total latency presented in  \cref{fig:intra-total_latency-performance}.
Roadrunner (User space) and Roadrunner (Kernel space) significantly reduce both total and serialization latency compared to RunC and WasmEdge, highlighting the impact of near-zero copy and serialization-free transfers on improving function communication performance.

\textbf{\underline{Inter-node:}} \textbf{\textit{Roadrunner} reduces the total latency by 62\% compared to WasmEdge and 7\% compared to RunC. For serialization latency, \textit{Roadrunner} reduces the serialization overhead by 97\% compared to WasmEdge and 46\% compared to RunC.} 

\begin{figure}[t]
\hspace*{-2em}
\centering
\begin{tikzpicture}
  \matrix (legend) [matrix of nodes,
                    nodes={inner sep=2pt, outer sep=0pt, font=\small},
                    column sep=6pt, row sep=0pt]
  {
    \tikz \node[fill=red!50,  draw=red,  minimum width=1em, minimum height=1em] {}; & Transfer      &
    \tikz \node[fill=blue!60, draw=blue, minimum width=1em, minimum height=1em] {}; & Serialization &
    \tikz \node[fill=cyan!60, draw=cyan, minimum width=1em, minimum height=1em] {}; & Wasm VM I/O   \\
  };
\end{tikzpicture}
\hspace*{-2em}
\begin{subfigure}[t]{0.3\linewidth}
    \hspace*{-3em}
    \begin{tikzpicture}
        \begin{axis}[
            ybar stacked,
            ymin=0,
            ymax=3,
            xtick=data,
            bar width=10pt,
            ylabel={Seconds},
            ymajorgrids=true,
            grid style=dashed,
            height=4cm,
            width=3.5cm,
            xticklabels from table={\testdatanew}{Label},
            xticklabel style={text width=2cm, align=right, anchor=east, yshift=-0.6em, xshift=0.9em}, 
            enlarge x limits=0.2 
        ]
            \addplot [fill=red!50]
                table [y=Transfer, meta=Label, x expr=\coordindex]
                    {\testdatanew};
            \addplot [fill=blue!60]
                table [y=Serialization, meta=Label, x expr=\coordindex]
                    {\testdatanew};
            \addplot [fill=cyan!60]
                table [y=WasmVM, meta=Label, x expr=\coordindex]
                    {\testdatanew};
        \end{axis}
    \end{tikzpicture}
    \centering
    \caption{}
    \label{fig:latency-fixed-size}
\end{subfigure}
\hspace{0em}
\begin{subfigure}[t]{0.3\linewidth}
    \hspace*{-3em}
    \begin{tikzpicture}
        \begin{axis} [
            box plot width=2mm,
            width=3.5cm,
            height=4cm,
            xtick={-1, 1.4, 3.7, 6},
            xticklabels={},
            ymajorgrids=true,
            grid style=dashed,
            enlarge x limits=0.2,
            ymode=log,
            xticklabels from table={\testdatanew}{Label},
            xticklabel style={text width=2cm, align=right, anchor=east, yshift=-0.6em, xshift=0.9em},
        ]
        \boxplot [orange, fill=orange!50] {serializationRoadrunner2.dat}
        \boxplot [color=blue, fill=blue!30] {serializationNative.dat}
        \boxplot [green!50!black, fill=green!85!black] {serializationWasmEdge.dat}
        \end{axis}
\end{tikzpicture}
    \caption{}
    \label{fig:serialization-free}
\end{subfigure}
\hspace{0em}
\begin{subfigure}[t]{0.3\linewidth}
    \hspace*{-2em}
    \begin{tikzpicture}
        \begin{axis}[
            ybar stacked,
            ymin=0,
            ymax=110,
            xtick=data,
            bar width=10pt,
            ylabel={Percentage (\%)},
            y label style={yshift=-0.2cm},
            yticklabel style={xshift=0.2em},
            ymajorgrids=true,
            grid style=dashed,
            height=4cm,
            width=3.5cm,
            xticklabels from table={\testdatanew}{Label},
            xticklabel style={text width=2cm, align=right, anchor=east, yshift=-0.6em, xshift=0.9em},
            enlarge x limits=0.2,
        ]
            \addplot [fill=red!50, nodes near coords, every node near coord/.append style={font=\scriptsize, black,rotate=90}]
                table [
                    y expr=\thisrow{Transfer} / (\thisrow{Transfer} + \thisrow{WasmVM} +\thisrow{Serialization}) * 100,
                    meta=Label, x expr=\coordindex
                ] {\testdatanew};
            \addplot [fill=blue!60]
                table [
                    y expr=\thisrow{Serialization} / (\thisrow{Transfer} + \thisrow{WasmVM} +\thisrow{Serialization}) * 100,
                    meta=Label, x expr=\coordindex
                ] {\testdatanew};
            \addplot [fill=cyan!60]
                table [
                    y expr=\thisrow{WasmVM} / (\thisrow{WasmVM} + \thisrow{Transfer} + \thisrow{Serialization}) * 100,
                    meta=Label, x expr=\coordindex
                ] {\testdatanew};
        \end{axis}
    \end{tikzpicture}
    \caption{}
    \label{fig:normalized-serialization-free}
\end{subfigure}
\caption{Breakdown of inter-node transfer latency for a 100MB payload across Roadrunner (RR), RunC (RC), and WasmEdge (W): (a) Detailed analysis of latency components, including serialization, communication transfer, and Wasm IO overhead; (b) Serialization overhead comparison, highlighting Roadrunner's near-zero copy and serialization-free mechanisms versus HTTP-based communication in RunC and WasmEdge; and (c) Normalized latency distribution, showcasing Roadrunner's efficiency in reducing serialization impact while maintaining competitive overall latency.}
\label{fig:comparison-latency}
\end{figure}
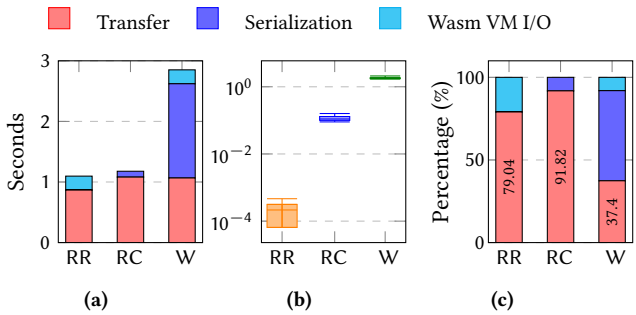

\cref{fig:comparison-latency} highlights the gains from Roadrunner compared to RunC and Wasmedge at a fixed size. \cref{fig:latency-fixed-size} shows the total latency contributions from data transfer, serialization, and Wasm VM IO overhead, where Wasm VM IO is the overhead Roadrunner takes to access the data from the host via the data access mechanisms presented in \cref{subsec:data_access}. Although Roadrunner presents gains in the transfer compared to RunC and Wasmedge, it pays a penalty (Wasm IO) to access the data in the Wasm VM. However, due to the near-zero copy and serialization-free improvements, shown in \cref{fig:serialization-free}, Roadrunner offers near-native speed to the Wasm-based functions, reducing the serialization overhead introduced by state-of-the-art communication mechanisms such as HTTP. In \cref{fig:normalized-serialization-free}, the normalized latency demonstrates that Roadrunner significantly reduces serialization overhead compared to RunC and WasmEdge. As a result, the overall latency in Roadrunner approaches that of RunC, where the majority of the latency is attributed to network transfer. As Roadrunner reduces overall latency, the relative share of Wasm IO overhead increases from 8\% in RunC to 20\% in Roadrunner.

\cref{fig:inter-overall_performance} shows the varying input size for Roadrunner and baselines. \cref{fig:inter-total_latency-performance} and \cref{fig:inter-total_throughput-performance} depict total latency and throughput, incorporating both data transfer and serialization overhead. Roadrunner demonstrates a reduction in total latency compared to RunC and WasmEdge, primarily due to its near-zero copy and serialization-free data transfer mechanism. However, the improvement margin is narrower in the inter-node scenario, as network transfer overheads become a dominant factor. \cref{fig:inter-serial_latency-performance} and \cref{fig:inter-serial_throughput-performance} focus on serialization overheads and reveal that Roadrunner significantly minimizes serialization latency. Although, in inter-node data transfer, the network contributes to the majority of the latency, Roadrunner offers Wasm-based serverless latency similar to RunC.

\pgfplotstableread{
Label Transfer Serialization   WasmVM
RR 0.866915105  0.000264914   0.229661058
RC 1.081091731 0.096302649  0 
W 1.066314933  1.555044937  0.229661058
    }\testdatanew

\input{plots/transfer_latency_different_payload}

\subsection{Scalability}

\input{plots/fanout}

\textbf{\underline{Intra-node:}} \textbf{\textit{Roadrunner (User space)} reduces the latency up to 70\% compared to RunC and increases the throughput up to 3x; reduces up to 98\% compared to Wasmedge and increases the throughput up to 64x. Roadrunner (Kernel space) reduces up to 77\% compared to Wasmedge and increases the throughput up to 4x.}

\cref{fig:intra-fanout} presents the performance metrics for intra-node fanout scenarios with 10MB function input transfers.
\cref{fig:intra-fanout-latency} and \cref{fig:intra-fanout-throughput} present the total latency and throughput across different fanout degrees for Roadrunner (User space), Roadrunner (Kernel space), RunC, and WasmEdge. Roadrunner (User space) demonstrates the lowest total latency due to its efficient near-zero copy mechanism. Roadrunner (Kernel space) exhibits slightly higher latency than RunC due to the overhead costs imposed by IPC in combination with async libraries to handle the fanout degrees. However, both significantly outperform WasmEdge, which suffers from high serialization costs and network transfer overhead. 

\cref{fig:intra-fanout-serial-latency} and \cref{fig:intra-fanout-serial-throughput} isolate the serialization impact, where Roadrunner (User space) and Roadrunner (Kernel space) maintain minimal serialization latency and high throughput, respectively. WasmEdge exhibits the highest serialization latency, which dominates its overall performance. RunC's serialization latency is higher than Roadrunner but still lower than Wasmedge.

\textbf{\underline{Inter-node:}} \textbf{\textit{Roadrunner} reduces the latency by up to 65\% compared to WasmEdge and increases the throughput up to 2.8x.}

\cref{fig:inter-fanout} shows scalability results increasing the fanout degree.
Similar as intra-node results, \cref{fig:inter-fanout-latency}
and \cref{fig:inter-fanout-throughput} depict the total latency and throughput across varying fanout degrees for Roadrunner, RunC, and WasmEdge. Roadrunner maintains a relatively consistent latency profile across fanout degrees similar to RunC and Wasmedge. Further, it shows a slight increase compared to RunC, indicating the IPC and async library overhead. Roadrunner outperforms WasmEdge, which shows significantly higher latencies due to substantial serialization costs. \cref{fig:inter-fanout-serial-latency} further isolates the serialization impact, emphasizing the advantages of Roadrunner near-zero copy and serialization-free mechanism. Meanwhile, \cref{fig:inter-fanout-serial-throughput} shows Roadrunner consistently outperforming WasmEdge and RunC achieving increased throughput.

\subsection{Resource Efficiency}

\textbf{\underline{Intra-node:}} \textbf{\textit{Roadrunner} reduces the CPU usage by up to 94\% compared to WasmEdge and reduces the RAM usage by up to 50\%}

During performance experiments that measure latency and throu\-ghput shown in \cref{fig:intra-different-payload}, Roadrunner exhibits significantly lower CPU usage compared to WasmEdge in \cref{fig:intra-cpu-performance}. More specifically, the lack of serialization and multiple copies between user and kernel space, Roadrunner decreases the user space CPU usage, shown in \cref{fig:intra-user-performance}. \cref{fig:intra-kernel-performance} shows Kernel Space CPU usage, where Roadrunner keeps consistently close to RunC, presenting a decrease compared to Wasmedge. 

During the scalability experiments, in \cref{fig:intra-fanout}, Roadrunner shows a constant CPU usage even with the fanout degree increase in \cref{fig:intra-fanout-cpu}. Although Roadrunner handles the data transfer alone in the user space, functions $a$ and $b$ are embedded in one single Wasm VM, which means it runs in a single process. In this case, the CPU handles both computation and data transfer, leading to an increased CPU burden, especially in the user space CPU usage in \cref{fig:intra-fanout-user}. Nevertheless, Roadrunner (User space) shows consistent and stable usage even for increased fanout degree. On the other hand, Roadrunner (Kernel space) IPC mechanisms result in a decreased user and kernel space CPU usage, shown respectively in \cref{fig:intra-fanout-user} and \cref{fig:intra-fanout-kernel}, resulting in an overall decrease in CPU usage (\cref{fig:intra-fanout-cpu}). The trend continues for RAM usage in \cref{fig:intra-fanout-ram} where the usage increases with the fanout degree and Roadrunner (Kernel space) outforms RunC and Wasmedge while Roadrunner (User space) shows a steady improvement compared to Wasmedge. 

\textbf{\underline{Inter-node:}} \textbf{\textit{Roadrunner} reduces the CPU usage by up to 85\% compared to WasmEdge and reduces the RAM usage by up to 25\%}

During performance experiments in \cref{fig:intra-different-payload}, 
Roadrunner shows a total CPU usage (\cref{fig:inter-performance-cpu}) overhead for small payload sizes, which is a consequence of an increased Kernel Space CPU Usage (\cref{fig:inter-performance-kernel}), but it stabilizes as the payload size increases, while Roadrunner user space CPU usage (\cref{fig:inter-performance-user}) keeps a linear increase similar to RunC and Wasmedge. This means Roadrunner mechanisms have a slight CPU usage overhead for small input sizes, whereas they show decreased CPU usage for larger input sizes. In \cref{fig:inter-performance-ram}, Roadrunner keeps RAM usage low for small payloads and increases linearly, similar to Wasmedge RAM consumption. Roadrunner accepts a higher CPU cost for small payloads to achieve high-speed communication for all payload sizes.

During scalability experiments in \cref{fig:inter-fanout}, Roadrunner shows a slightly increased CPU usage (\cref{fig:inter-fanout-cpu}) for small payloads, but then it stabilizes and keeps a slight linear increase over the fanout increase. A similar trend can be noticed in \cref{fig:inter-performance-cpu}. Under high load, Roadrunner reports less User CPU usage (\cref{fig:inter-fanout-user}) than Wasmedge and RunC, showing that although the near-zero copy and serialization-free mechanisms might incur slightly higher CPU usage for smaller input sizes, they incur less CPU usage for larger input sizes due to the reduced copies between user space and kernel space. In \cref{fig:inter-fanout-kernel}, the increase is linear for Roadrunner and baselines. In \cref{fig:inter-fanout-ram}, Roadrunner shows a decreased RAM usage for small input sizes but higher RAM usage as the fanout degree increases. This increase is attributed to the additional memory requirements for handling network buffers, which do not exist for intra-node data transfer since Roadrunner leverages IPC to exchange data.

\section{Discussion}\label{sec:discussion}

\paragraph{Benefits and Trade-Offs} Each Roadrunner communication mechanism suits the demands of different workflows. Selecting the most appropriate communication depends on balancing the benefits and trade-offs between communication frequency, data load, latency, scalability, and resource utilization. Therefore, it is crucial to identify when and which communication is most appropriate for a given workflow, ensuring that the selected mode also aligns with the workflow's defined requirements. User space data transfer ensures low latency and minimal resource contention but is tightly coupled, limiting scalability and introducing fault propagation risks. Kernel space data transfer provides isolation and independent scaling via IPC mechanisms, such as shared memory or UNIX domain sockets, but incurs IPC overhead and increases application complexity. Network transfer across hosts enables high scalability and strong isolation but suffers from increased latency due to network overhead alongside operational costs. 

\paragraph{Interoperability} Roadrunner's data transfer model ensures seamless inter-function communication, eliminating the dependency on external storage services for data exchange. This design is not limited to inter-function communication; it can be extended to scenarios involving external services with large input data, reducing the additional overhead associated with serialization, which significantly affects Wasm-based serverless functions. Moreover, Roadrunner remains backward compatible by defaulting to standard WASI communication unless its API is explicitly invoked, allowing unmodified modules to operate normally while enabling near-zero copy, serialization-free communication for those that opt for Roadrunner's inter-function communication.

\paragraph{Near-zero Copy Data Transfer}
Roadrunner implements near-zero copy data transfer by using zero-copy syscalls such as \texttt{splice} and \texttt{vmsplice} to move data directly between the shim and kernel buffers, avoiding redundant copies within the kernel. While this eliminates overhead between user and kernel space, data must still be copied in and out of the Wasm VM's linear memory due to Wasm's isolation model.  Therefore, we adopted the term ``near-zero copy''.

\paragraph{Security Concerns} The Wasm VM is designed to ensure strong isolation between the host and the Wasm VM, preventing unauthorized access or manipulation of memory. Although the host accesses the Wasm modules' linear memory regions, risks such as memory corruption, buffer overflow, and data leakage are avoided due to the strict memory boundaries from Wasm which means, that a specific Wasm VM memory region is only accessible to the specific process that spawns the Wasm VM, i.e., the function shim. Every Wasm module has completely isolated memory regions not accessible to each other. 
In the event of a boundary violation, the function execution simply fails without affecting other parts of the system. Additionally, the memory access APIs, described in \cref{tab:functions}, are designed to enforce strict access controls, preventing malicious or compromised host processes from exploiting the direct memory interface.

\paragraph{Threats to Validity} To the best of our knowledge, Roadrunner is the first to explore serialization-free data transfer for Wasm serverless functions, which introduces certain limitations in the evaluation scope.  Consequently, our evaluation focuses on comparing Roadrunner against state-of-the-art solutions used in serverless platforms that rely on traditional communication methods, such as HTTP, which rely on IO serialization.
Although our evaluation presents the most common invocation patterns~\cite{CloudProgrammingSimplified}, it is conducted in controlled environments with specific workloads, which may not fully capture the diversity of real-world serverless applications. 
Additionally, the performance gains observed may vary depending on the characteristics of the underlying hardware, network conditions, and application patterns. Roadrunner reliance on underlying host mechanisms may limit its adoption, however the demonstrated improvements highlights its potential as a foundation for future Wasm-native serverless platforms.







\section{Related Work}
\label{sec:rel_work}

This section provides an overview of existing research that addresses Wasm-based Serverless Data Transfer between Serverless Functions and Sandoxes and Shims, highlighting their contributions, limitations, and how they compare to Roadrunner.

\subsection{WebAssembly-based Serverless}
Faasm ~\cite{faasm} proposes a stateful Wasm-based serverless platform. To achieve that Faasm introduces Faaslet, a lightweight sandbox isolation mechanism that relies on cgroups and Wasm isolation to execute serverless functions. Moreover, Faasm relies on Faabric ~\cite{faabric} to manage and exchange data between serverless functions. However, Faasm introduces a dedicated serverless platform not compatible with state-of-the-art platforms. Sledge~\cite{sledge} introduces a serverless Wasm runtime designed for edge computing, focusing on data-intensive multi-tenancy. It leverages Wasm sandboxes for lightweight isolation and introduces its own scheduler bypassing the kernel to optimize the execution performance and reduce resource overhead.  Nevertheless, Sledge relies on a custom Wasm runtime, which limits compatibility with state-of-the-art serverless and increases development and maintenance complexity.
CWASI~\cite{Cwasi2023} proposes an OCI-compliant shim that leverages function locality to enable direct inter-function communication: Function Embedding, Local Buffer, and Networked buffer. CWASI Function Embedding enables high-speed data passing, such as shared memory. Local buffer enables inter-function communication via host mechanisms.  Nevertheless, CWASI leverages remote storage services for functions placed on different hosts, increasing latency and network overhead. 
Wow~\cite{PushingWasm} leverages Wasm to provide isolation and portability for heterogeneous environments such as the Edge-Cloud continuum. Wow enables serverless workflows with decreased cold start times. Nevertheless, Wow does not address communication or data transfer, which accounts for a big portion of the serverless function execution time. PSL~\cite{PSL2024} leverages Wasm's deny-by-default and isolation mechanisms to execute security and privacy in serverless workloads. PSL relies on a state store to manage state and communication among functions. Nevertheless, it still relies on state-of-the-art communication mechanisms such as HTTP, which incurs serialization and, consequently, latency increase and resource usage overhead. 
Roadrunner achieves serialization-free communication by leveraging Wasm's linear memory model to facilitate data exchange between functions. Unlike traditional methods that rely on serialization to encode data for transmission (e.g., via HTTP), Roadrunner maps and transmits memory regions within the function via kernel buffers, minimizing the overhead associated with data preparation, serialization, and deserialization while maintaining the integrity and isolation of the Wasm sandbox.

\subsection{Serverless Inter-function Communication}

SAND~\cite{sand} introduces a hierarchical message bus to provide low latency with a global and a local bus. SAND optimizes remote inter-function communication by introducing a global message bus responsible for exchanging data between functions remotely. 
Moreover, InfiniCache ~\cite{InfiniCache} enables in-memory cache communication between serverless functions by providing the proxy address attached to each function.
Boxer~\cite{boxer} and FMI~\cite{fmi} explore NAT TCP punch hole. The punch-hole network technique enables direct communication between two clients. In this approach, the rendezvous server controls and exchanges address data between two clients. 
For example, function A requests a connection to function B to the server S, the server S replies to function A and function B addresses, and forwards function A addresses to function B. Once both functions are aware of each other, they can exchange data directly.

\paragraph{Zero-copy} Diffuse~\cite{diffuse} presents DSMQueue, a distributed shared-memory queue. DSMQueue leverages RDMA techniques to enable zero-copy data exchange between serverless functions. Nevertheless, it relies on specialized hardware, which might be a bottleneck in resource-constrained environments such as the Edge-Cloud continuum. In ~\cite{following-data}, authors propose shared memory via co-located function data exchange and a KVS for remote communication.
Nightcore~\cite{nightcore} places functions from the same workflow in a single host to enable low-latency communication via shared memory. However, nightcore does not address data serialization, which comprises a large portion of the function communication. 
Wasmer~\cite{wasmer42} reduces cold-start latency by enabling zero-copy deserialization of precompiled Wasm modules through memory-mapped metadata access. It serializes metadata and compiled code into a cache file that the runtime can later map directly into memory, avoiding parsing, reconstruction, and copying overhead. However, Wasmer does not support zero-copy for runtime inputs, which are still passed via traditional memory copies and host interfaces. While Wasmer focuses on cold starts, RoadRunner eliminates serialization overhead in data exchange between functions during runtime, also enabling direct communication. 
SharedArrayBuffer~\cite{MDN_SharedArrayBuffer}  enables direct memory sharing between workers, which means it enables memory sharing between workers and threads within the same function, but it does not allow between multiple processes. Roadrunner enables data sharing at the inter-function level. Furthermore, Roadrunner takes a stricter approach to memory isolation than SharedArrayBuffer, all memory sharing in Roadrunner is explicitly mediated by the shim, which controls allocation, visibility, and access semantics.

\paragraph{Serialization-free} RMMap~\cite{Serialization2024} introduces an OS primitive for remote memory map which enables serverless functions to directly access the memory of another function. Nevertheless, RMMap is designed for RDMA-based network transfer, which requires specialized hardware. PraaS~\cite{copik2024praas} enables functions running as processes, thus co-located functions achieve serialization-free communication by using a shared memory pool for direct object access, allowing zero-copy exchange via pointers in languages such as C, while languages such as Python still use pickling but with less overhead than traditional IPC.

Roadrunner enables inter-function communication, leveraging host mechanisms to exchange data and decrease serialization overhead. Moreover, Roadrunner introduces near-zero copy techniques that optimize the communication latency and throughput of data exchange between serverless functions.

\section{Conclusion and Future Work}
\label{sec:con}

In this paper, we introduced Roadrunner, a near-zero copy, and serialization-free data transfer serverless runtime shim designed to optimize inter-function communication in WebAssembly-based serverless functions. Roadrunner leverages Wasm's linear memory model and syscalls to minimize data transfer overhead between user and kernel space. By eliminating unnecessary serialization and reducing latency, Roadrunner significantly enhances the performance of serverless functions, particularly in data-intensive and latency-sensitive applications.

We evaluated Roadrunner using serverless functions that represent common invocation patterns, such as sequence and parallel functions. Our experiments demonstrated that Roadrunner reduces data transfer latency and improves throughput compared to the data-passing mechanisms of state-of-the-art serverless platforms. The experiments show that Roadrunner improves latency by up to 89\% and increases the throughput up to 69x compared to the Wasm baselines, offering close to native speed for IO-bound serverless workloads.
These results highlight Roadrunner's capability to address the key bottlenecks of inter-function communication in serverless computing.

In the future, we plan to expand Roadrunner beyond its current implementation to support a broader range of WebAssembly runtimes, enabling greater portability and integration across diverse serverless platforms. Our future work will focus on developing Roadrunner into a dynamic virtualization runtime that can autonomously select the runtime type, e.g., container and Wasm, and select the most suitable runtime for specific serverless workflows based on workload and environment characteristics. Moreover, we plan to extend Roadrunner with a self-provisioning infrastructure~\cite{2024selfprovisioningInfrastructure} that automatically provides the necessary BaaS for each function. Finally, we aim to introduce function state management and syscall batching, allowing Roadrunner to efficiently handle stateless and stateful serverless functions.

\section*{Acknowledgment}
This work is partially funded by the Austrian Research Promotion Agency (FFG) under the project RapidREC (Project No. 903884).
This work has received funding from the Austrian Internet Stiftung under the NetIdee project LEO Trek (ID~7442).
Partly funded by EU Horizon Europe under GA no. 101070186 (TEADAL) and Ga no. 101192912 (NexaSphere).

\bibliographystyle{ACM-Reference-Format}
\bibliography{base}

\balance    


\end{document}